\documentclass[final,times,5p,sort&compress]{elsarticle}

\usepackage{lineno,hyperref}

\usepackage{amsmath}
\usepackage{amssymb}
\usepackage{graphicx}
\usepackage{bm}
\usepackage{braket}
\usepackage{mathrsfs}
\usepackage{accents}
\usepackage{longtable}
\usepackage{rotating}
\usepackage{cancel}

\modulolinenumbers[5]

\journal{Physics Letters B}

\begin{document}

\begin{frontmatter}

\title{
Spin entanglement of multinucleons: experimental prospects
}

\cortext[mycorrespondingauthor]{Corresponding author}

\author[hhu,fd]{Dong Bai\corref{mycorrespondingauthor}}
\ead{dbai@hhu.edu.cn}

\author[tj1,tj2]{Zhongzhou Ren\corref{mycorrespondingauthor}}
\ead{zren@tongji.edu.cn}

\address[hhu]{College of Mechanics and Engineering Science, Hohai University, Nanjing 211100, China}
\address[fd]{Shanghai Research Center for Theoretical Nuclear Physics, NSFC and Fudan University, Shanghai 200438, China}
\address[tj1]{School of Physics Science and Engineering, Tongji University, Shanghai 200092, China}
\address[tj2]{Key Laboratory of Advanced Micro-Structure Materials, Ministry of Education, Shanghai 200092, China}

\begin{abstract}

Multiprotons and multineutrons are among the most exotic and mysterious things ever produced on earth.
They provide an exceptional opportunity to understand nuclear forces and nuclear dynamics at extreme conditions, as well as neutron stars in the heaven.
Quantum entanglement, referred to as ``spooky action at a distance'' by Einstein, is a ubiquitous yet deep property of quantum systems.
It not only occupies a central position in quantum information science but also is investigated intensively in high energy physics, condensed matter physics, and quantum gravity. 
In comparison, the study of nuclear entanglement is still in infancy, and the entanglement properties of multiprotons and multineutrons in free space are generally unknown.
Here, we study the crucial problem of how to measure spin entanglement of these multinucleons in nuclear experiments,
with
special emphases on two- and three-nucleon states.
These findings open a freshly new direction for the multinucleon research.
They are also useful for understanding entanglement properties of other exotic nuclear objects.

\end{abstract}


\date{}%

\end{frontmatter}


\section{Introduction}

Multiprotons and multineutrons
are among the most exotic objects in nuclear physics. 
The nature of these multinucleon systems remains appealing and mysterious to physicists
despite of in-depth studies lasting for decades.\footnote{Here, the word ``multinucleon'' is used as the general term for the multiproton and the multineutron in free space.
Therefore, the other multinucleons like ${}^4$He are not under consideration.
}
There has been a long and unsettled debate since the 1960s on the existence of bound or resonant multineutron systems in free space \cite{Marques:2021mqf}.
Recently, a correlated free multineutron system with $N=4$ has been observed,
with the extracted energy and width agreeing well with \emph{ab initio} results in favor of a multineutron resonance \cite{Duer:2022}.
However, 
it is found that 
the same experimental data could also be explained without making use of multineutron resonances \cite{Lazauskas:2022mvq}.
Multineutrons are also testing grounds for new concepts and tools, often leading to unexpected discoveries.
In Ref.~\cite{Hammer:2021zxb}, 
it is proposed that multineutrons can be identified, under certain conditions, as novel objects called ``unnuclei'', obeying nonrelativistic conformal symmetry instead of merely the Galilean symmetry.
On the other hand, multiprotons, the charged counterpart of multineutrons, appear naturally as decay products of many unstable proton-rich nuclei \cite{Pfutzner:2011ju,Zhou:2022yzf,Pfutzner:2023tvr}.
They provide valuable information of nuclear physics in extremely proton-rich environments.

On the other hand, the past three decades have witnessed huge progress in quantum information science (QIS),
which makes heavy use of quantum entanglement, a ubiquitous yet deep property of quantum systems \cite{Nielsen:2010,Benenti:2019}. 
Quantum entanglement is widely recognized as the key factor that makes quantum mechanics different
and lies behind almost every high-perform- ance protocol and algorithm in QIS.
It has been explored thoroughly in high energy physics \cite{Barr:2024djo}, condensed matter physics \cite{Laflorencie:2015eck}, and quantum gravity \cite{Rangamani:2016dms}.
In comparison, 
the interdisciplinary direction of nuclear entanglement is still in infancy \cite{Lamehi-Rachti:1976,Sakai:2006,Kwasniewicz:2013cqa,Kanada-Enyo:2015ncq,Kanada-Enyo:2015kyo,Legeza:2015fja,Gorton:2018,Johnson:2019,Beane:2018oxh,Kruppa:2020rfa,Robin:2020aeh,Low:2021ufv,Beane:2021zvo,Jafarizadeh:2022kcq,Kovacs:2021yme,Pazy:2022mmg,Tichai:2022bxr,Bai:2022hfv,Liu:2022grf,Johnson:2022mzk,Bai:2023rkc,Gu:2023aoc,Bulgac:2022cjg,Bulgac:2022ygo,Bai:2023tey,Miller:2023ujx,Miller:2023snw,Hengstenberg:2023ryt,Perez-Obiol:2023wdz,Bai:2023hrz,Kirchner:2023dvg},
and the entanglement properties of multinucleons remain to be discovered.
%

In this Letter, we focus on the key problem of experimental measurement of spin entanglement of multiprotons and multineutrons.
Here, spin entanglement is a specific kind of quantum entanglement associated with the spin wave function,
 with the 
 nucleon spin identified as a qubit, the basic unit of quantum information processing.
It has good mathematical properties compared to quantum entanglement of the full wave function,
 and has been shown to be intimately related to symmetry emergence of strong interaction at low energies \cite{Beane:2018oxh,Bai:2023tey,Bai:2022hfv,Liu:2022grf,Bai:2023rkc}.
In Ref.~\cite{Bai:2023hrz}, 
we study the problem of how to determine spin entanglement of the proton-neutron pair experimentally.
This Letter generalizes our previous work to the multiprotons and multineutrons,
with special emphases on the cases of two and three nucleons.
Compared with Ref.~\cite{Bai:2023hrz}, our main innovations are twofold.
First, the discussions are extended from two nucleons to three nucleons. 
Second, the discussions are extended from distinguishable particles to indistinguishable particles.
These extensions are crucial for experimental and theoretical studies of multiproton and multineutron spin entanglement.

The rest parts are organized as follows: In Sec.\ \ref{EM},
the essential background of quantum entanglement is introduced,
including concurrence and its generalizations to the three-qubit system.
In Sec.\ \ref{ExpMea}, some general considerations are presented first for measuring spin entanglement of multinucleons,
followed by nuclear state tomography (NST),
which is the nuclear physics realization of quantum state tomography developed in quantum information science \cite{Altepeter:2005} and is aimed at providing the key building blocks of entanglement measures.
In Sec.\ \ref{Concl},
conclusions and outlooks are given.

\section{Entanglement Measures}
\label{EM}

\subsection{Concurrence}
\label{TwQS}

Quantum entanglement highlights the existence of nonlocal correlations shared by different quantum objects.
%
Consider a general two-qubit state
$\ket{\Psi_{12}}=b_{00}\ket{0_10_2}+b_{01}\ket{0_11_2}+b_{10}\ket{1_10_2}+b_{11}\ket{1_11_2}$.
Here, $\ket{k_1l_2}$ abbreviates the tensor product state $\ket{k_1}\otimes\ket{l_2}$, with $k,l=0,1$ and $\ket{0_i}$ and $\ket{1_i}$ being the basis states for the $i$th qubit,
and $b_{kl}$ is the corresponding probability amplitude.
The subscripts ``12'' in $\ket{\Psi_{12}}$ stress that this is a two-qubit state for Qubits 1 and 2.
If $\ket{\Psi_{12}}$ can be rewritten as
$\ket{\Psi_{12}}=\ket{\alpha_1}\otimes\ket{\beta_2}$,
then it is a product state with no entanglement.
Here, $\ket{\alpha_1}=\tilde{\alpha}_0 \ket{0_1}+\tilde{\alpha}_1\ket{1_1}$ and $\ket{\beta_1}=\tilde{\beta}_0 \ket{0_2}+\tilde{\beta}_1\ket{1_2}$,
with $\tilde{\alpha}_0$ and $\tilde{\alpha}_1$ being the probability amplitudes of $\ket{\alpha_1}$,
and $\tilde{\beta}_0$ and $\tilde{\beta}_1$ being the probability amplitudes of $\ket{\beta_2}$.
In this case, one has
$b_{00}=\tilde{\alpha}_0\tilde{\beta}_0$, $b_{01}=\tilde{\alpha}_0\tilde{\beta}_1$, $b_{10}=\tilde{\alpha}_1\tilde{\beta}_0$, $b_{11}=\tilde{\alpha}_1\tilde{\beta}_1$,
such that $b_{00}b_{11}-b_{01}b_{10}=0$.
It is natural to expect that $\ket{\Psi_{12}}$ is entangled if 
\begin{align}
\mathcal{C}_{1|2}(\ket{\Psi_{12}})\equiv2 |b_{00}b_{11}-b_{01}b_{10}|\neq0.
\label{concurrence}
\end{align}
This is indeed the case as shown by Ref.~\cite{Hill:1997}, and $\mathcal{C}_{1|2}(\ket{\Psi_{12}})$ is known as the concurrence of $\ket{\Psi_{12}}$. Furthermore, $\mathcal{C}_{1|2}(\ket{\Psi_{12}})$ takes its value from 0 to 1 if $\ket{\Psi_{12}}$ is normalized.
If $\mathcal{C}_{1|2}(\ket{\Psi_{12}})$ equals 0, the two-qubit state $\ket{\Psi_{12}}$ is not entangled.
Otherwise, the larger $\mathcal{C}_{1|2}(\ket{\Psi_{12}})$ is, the more entangled $\ket{\Psi_{12}}$ will be.
When $\mathcal{C}_{1|2}(\ket{\Psi_{12}})$ equals 1, $\ket{\Psi_{12}}$ is called maximally entangled. 
For later convenience, it is helpful to introduce an equivalent definition of the concurrence
\begin{align}
\mathcal{C}_{1|2}(\ket{\Psi_{12}})=\sqrt{2\left[1-\text{tr}(\rho_1^2)\right]},
\label{concurrence2}
\end{align}
where $\rho_1=\text{tr}_2(\ket{\Psi_{12}}\bra{\Psi_{12}})$ is the reduced density matrix after tracing out the 2nd part of the bipartite state.
This definition also allows the notion of the concurrence to be generalized to bipartite states not made of two qubits.
Using the Schmidt decomposition, one can easily show that the concurrence is symmetric with respect to the bipartition in the sense that $\mathcal{C}_{1|2}(\ket{\Psi_{12}})=\mathcal{C}_{2|1}(\ket{\Psi_{12}})\equiv\sqrt{2\left[1-\text{tr}(\rho_2^2)\right]}$.

\subsection{Genuine multipartite entanglement measures}
\label{TQS}

Entanglement properties of three-qubit states are more complicated than two-qubit states.
According to Ref.~\cite{Ma:2011yon}, three-qubit states can be classified into three classes: product states, biseparable states, and genuine tripartite entangled states. A three-qubit state $\ket{\Psi_{123}}$ is called a product state if it could be put into
$\ket{\Psi_{123}}=\ket{\alpha_1}\otimes\ket{\beta_2}\otimes\ket{\gamma_3}$,
where $\ket{\alpha_1}$ and $\ket{\beta_2}$ follows the definitions in Sec.\ \ref{TwQS},
and $\ket{\gamma_3}$ is given by
$\ket{\gamma_3}=\tilde{\gamma}_0\ket{0_3}+\tilde{\gamma}_1\ket{1_3}$,
with $\tilde{\gamma}_0$ and $\tilde{\gamma}_1$ being the corresponding probability amplitudes.
The subscripts ``123'' in $\ket{\Psi_{123}}$ stress that this is a three-qubit state for Qubits 1, 2, and 3.
$\ket{\Psi_{123}}$ is called a biseparable state if it could be put into one of the three forms
$\ket{\Psi_{123}}=\ket{\alpha_1}\otimes\ket{\Psi_{23}}$,
$\ket{\Psi_{123}}=\ket{\beta_2}\otimes\ket{\Psi_{13}}$,
and
$\ket{\Psi_{123}}=\ket{\Psi_{12}}\otimes\ket{\gamma_3}$.
Otherwise, $\ket{\Psi_{123}}$ is called a genuine tripartite entangled state.

In Ref.~\cite{Ma:2011yon}, it is proposed that a good entanglement measure for three-qubit states should be zero for all product and biseparable states, and positive for all genuine entangled states.
It turns out that only a few entanglement measures satisfy the above requirements, and they are referred to as genuine multipartite entanglement (GME) measures. Three of them are genuine multipartite concurrence (GMC) \cite{Ma:2011yon}, geometric mean of bipartite concurrences (GBC) \cite{Li:2021hhj}, and concurrence fill (CF) \cite{Xie:2021hsy},
\begin{align}
&\mathcal{C}_\text{GMC}(\ket{\Psi_{123}})=\text{min}\{\mathcal{C}_{1|23},\mathcal{C}_{2|13},\mathcal{C}_{3|12}\},
\label{CGMC3}\\
&\mathcal{C}_\text{GBC}(\ket{\Psi_{123}})=\left[\mathcal{C}_{1|23}\mathcal{C}_{2|13}\mathcal{C}_{3|12}\right]^{\tfrac{1}{3}},\\
&\mathcal{C}_\text{CF}(\ket{\Psi_{123}})=\left[\frac{16}{3}\mathcal{Q}(\mathcal{Q}-\mathcal{C}_{1|23}^2)(\mathcal{Q}-\mathcal{C}_{2|13}^2)(\mathcal{Q}-\mathcal{C}_{3|12}^2)\right]^{\tfrac{1}{4}},\nonumber\\
&\text{with}\ \mathcal{Q}=\frac{1}{2}\left(\mathcal{C}^2_{1|23}+\mathcal{C}^2_{2|13}+\mathcal{C}^2_{3|12}\right).
\label{CCF3}
\end{align}
Here, $\mathcal{C}_{1|23}$ is the bipartite concurrence of $\ket{\Psi_{123}}$ between Qubit 1 and Qubits $(23)$,
which is given by Eq.~\eqref{concurrence2} with $\rho_1$ reinterpreted as the reduced density matrix after tracing out the 2nd and 3rd qubits,
and
$\mathcal{C}_{2|13}$ and $\mathcal{C}_{3|12}$ are defined in a similar way.
It is straightforward to see that the bipartite concurrences for different bipartitions of $\ket{\Psi_{123}}$ play a fundamental role in defining GME measures for three-qubit systems.
There are other GME measures in the literature, and their total number keeps growing.
These GME measures tend to give different values for the amount of entanglement,
and it is not fully clear whether 
they are guaranteed to give the same ranking for three-qubit states.
Therefore, it is helpful to consider different GME measures simultaneously.

 \begin{figure}

\centering
  \includegraphics[width=0.6\linewidth]{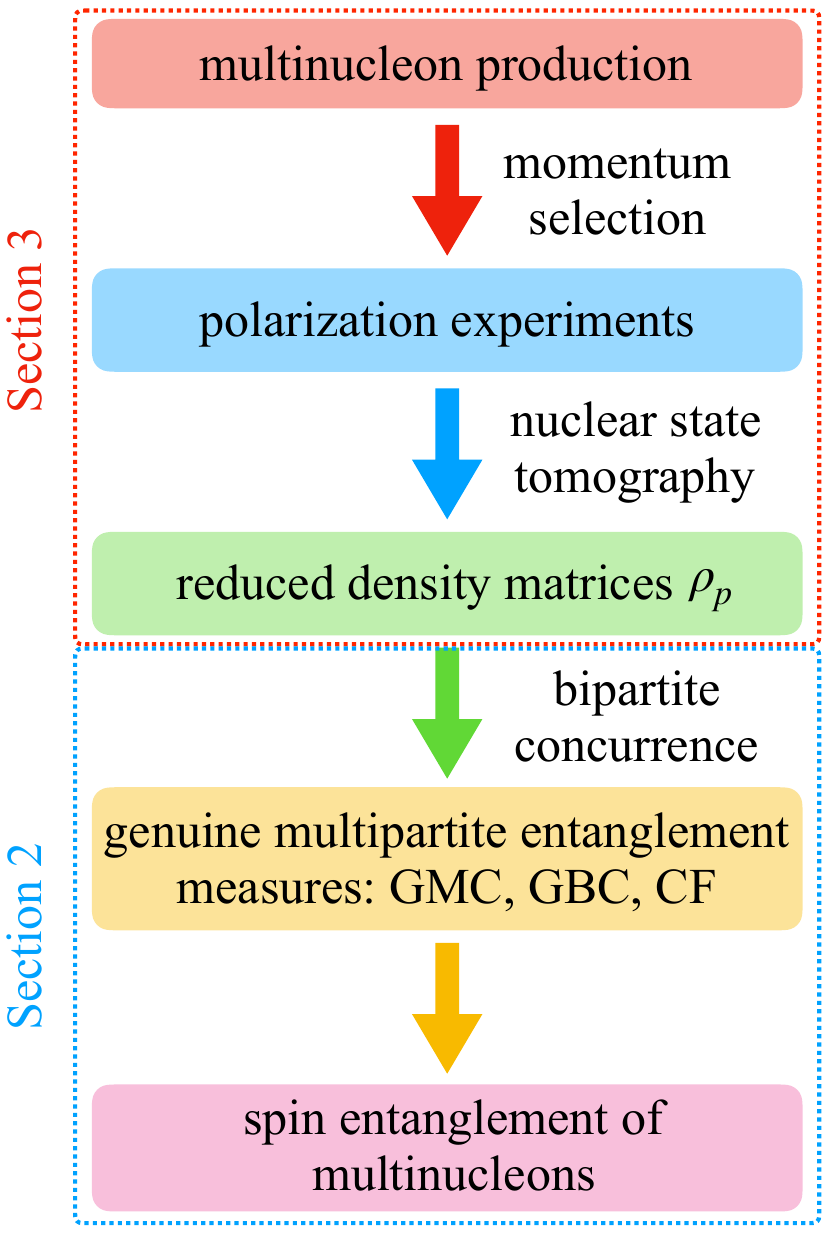}
  \caption{The flow chart for determining spin entanglement of multinucleons experimentally.
  The multinucleons are produced first via nuclear decays and nuclear reactions, followed by momentum selection.
  The one- and two-nucleon reduced density matrices $\rho_p$ ($p=1,2$ for the two-nucleon state and $p=1,2,3$ for the three-nucleon state)
  are obtained by nuclear state tomography (NST) from polarization experiments.
  The genuine multipartite entanglement (GME) measures, which quantify the spin entanglement of multinucleons, can then be derived from the reduced density matrices $\rho_p$ via bipartite concurrences.
  }

  \label{streamline}
       
\end{figure}


\subsection{Mode Entanglement}

In the above discussions, 
it is assumed implicitly that all the qubits are realized by distinguishable particles that can be addressed individually.
Therefore, they cannot be applied to multiprotons and multineutrons directly as the latter are made of indistinguishable particles.
In Ref.~\cite{Benatti:2020}, five popular definitions of entanglement for indistinguishable particles are examined under three consistency criteria:
no entanglement should be generated by local operators, the standard notion of entanglement should be recovered after addressing all the particles, and no information processing can outperform the classical ones by using local operators only.
It is concluded that there is only one definition satisfies all three criteria, i.e., mode entanglement. 

In second quantization, a nucleon with the momentum $\bm{k}$ and the helicity $\lambda=\pm\tfrac{1}{2}$ is given by
$\ket{\bm{k}\lambda}=a^\dagger({\bm{k}\lambda})\ket{\Omega}$,
with $a^\dagger({\bm{k}\lambda})$ being the mode creation operator satisfying the anticommutation relation $\{a(\bm{k}\lambda),a^\dagger(\bm{k}'\lambda')\}$ $=\delta^3(\bm{k}-\bm{k}')\delta_{\lambda\lambda'}$ and $\ket{\Omega}$ being the vacuum state.
$\ket{\bm{k}\lambda}$ satisfies the normalization condition $\braket{\bm{k}'\lambda'|\bm{k}\lambda}=\delta^3(\bm{k}-\bm{k}')\delta_{\lambda\lambda'}$.
When $\bm{k}$ is fixed (``frozen''), $\ket{\bm{k}\lambda}$ could be regarded as a qubit with $\ket{\bm{k}\frac{1}{2}}\to\ket{0}$
and $\ket{\bm{k},-\frac{1}{2}}\to\ket{1}$.
The quantum state for two nucleons with momenta $\bm{k}_1$, $\bm{k}_2$ and helicities $\lambda_1$, $\lambda_2$ is given by
$\ket{\bm{k}_1\lambda_1,\bm{k}_2\lambda_2}=a^\dagger(\bm{k}_1\lambda_1)\,a^\dagger(\bm{k}_2\lambda_2)\ket{\Omega}$.
In mode entanglement,
$\ket{\bm{k}_1\lambda_1,\bm{k}_2\lambda_2}$ is a product state with no entanglement, despite that
it corresponds to a Slater determinant in first quantization and may look as if it is entangled.
A general two-nucleon state with frozen momenta $\bm{k}_1$ and $\bm{k}_2$ is then mapped to a two-qubit state
$\ket{\Psi_{\bm{k}_1\bm{k}_2}}\equiv\sum_{\lambda_1\lambda_2}\mathcal{B}(\bm{k}_1\lambda_1,\bm{k}_2\lambda_2)$ $\times\ket{\bm{k}_1\lambda_1,\bm{k}_2\lambda_2}
\to\sum_{\lambda_1\lambda_2}\mathcal{B}(\bm{k}_1\lambda_1,\bm{k}_2\lambda_2)\ket{\tfrac{1}{2}-\lambda_1,\tfrac{1}{2}-\lambda_2}$,
and the three-nucleon states with frozen momenta can be mapped to three-qubit states in a similar way.
The entanglement properties can then be quantified by following previous subsections.
It is easy to check that mode entanglement is reduced to the standard notion in the case of distinguishable particles.

%
%

\section{Experimental Prospects of Measuring Spin Entanglement}
\label{ExpMea}

\subsection{General Considerations}

\begin{figure}

\centering
  \includegraphics[width=\linewidth]{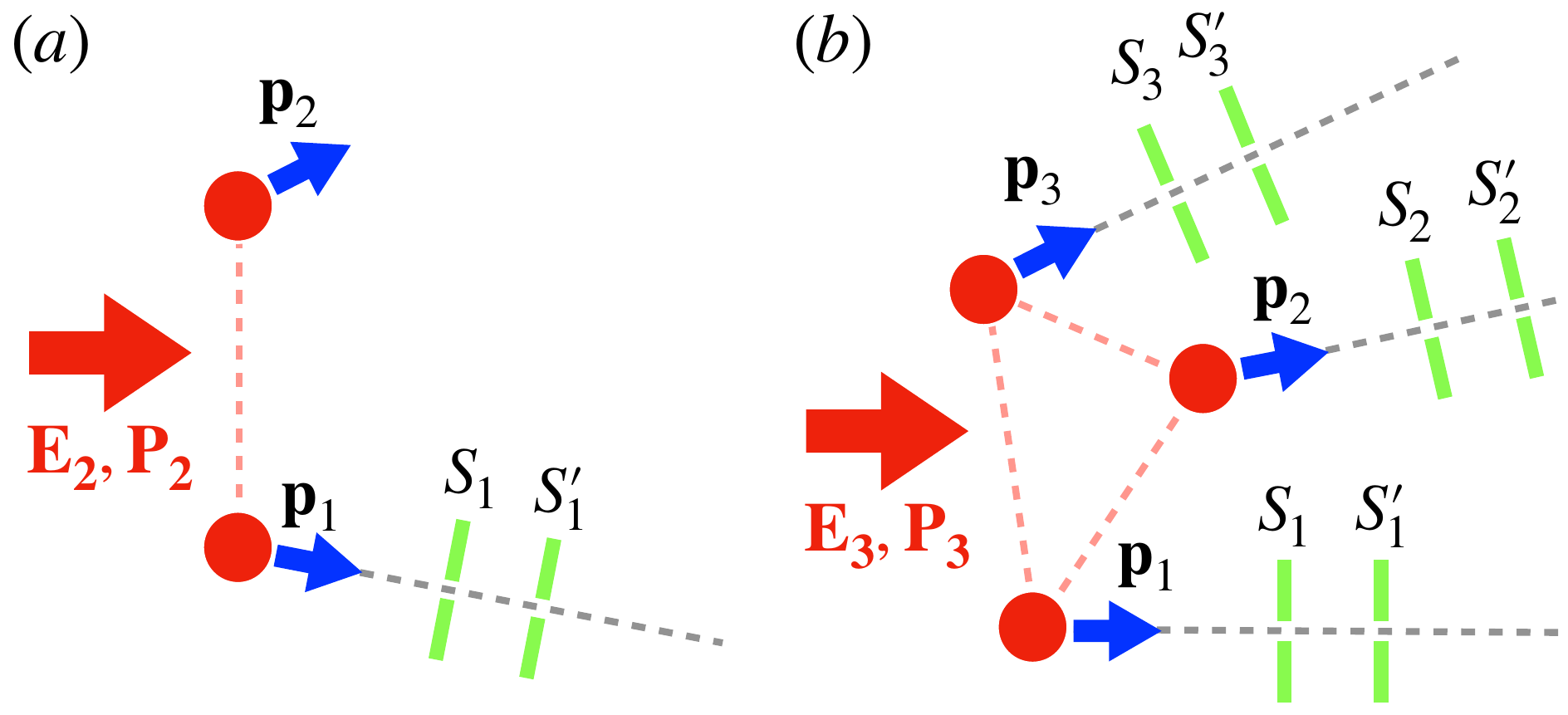}
  \caption{
  A proof-of-concept experimental setup for fixing the nucleon momenta of (a) the two-proton or two-neutron state and (b) the three-proton or three-neutron state.
  The pink dashed lines between nucleons highlight the spin entanglement shared by the nucleons.
  The total energies and the total momenta of multinucleons are colored in red,
  while the momenta of individual nucleons are colored in blue.
  The green rectangles are narrow slits (namely, $S_i$ and $S_i'$, with $i=1,2,3$) which select outgoing nucleons in specific directions.
  In Figure \ref{TMF}(b), the three nucleons should all pass the slits. In order to reduce the impacts from the imperfect events where only one or two nucleons pass the slits, $S_i$ and $S_i'$ could actually be made of nucleon detectors that record the nucleon hitting the slits.
  This may be helpful for the post-selection of legitimate events.
}

  \label{TMF}
       
\end{figure}

The general procedure to measure spin entanglement of multinucleons is given in Fig.~\ref{streamline}.
To measure spin entanglement experimentally, it is required that a definite multinucleon wave function can be produced for a large number of times.
A general wave function for multinucleons can be given in momentum space by
\begin{align}
\ket{\overline{\Phi}_{1\cdots N}}=&\sum_{\overline{\lambda}_1\cdots\overline{\lambda}_N}\!\int\!\mathrm{d}^3 k_1\cdots\!\int\!\mathrm{d}^3 k_N\ 
\mathcal{B}(\bm{k}_1\overline{\lambda}_1,\cdots,\bm{k}_N\overline{\lambda}_N)\nonumber\\
\times&\ket{\bm{k}_1\overline{\lambda}_1,\cdots,\bm{k}_N\overline{\lambda}_N}.
\end{align}
Thanks to the Pauli principle, the probability amplitude $\mathcal{B}(\bm{k}_1\overline{\lambda}_1,$ $\cdots,\bm{k}_N\overline{\lambda}_N)$
obeys $\mathcal{B}(\cdots, \bm{k}_i\overline{\lambda}_i, \cdots,\bm{k}_j\overline{\lambda}_j,\cdots)=-\mathcal{B}(\cdots, \bm{k}_j\overline{\lambda}_j,$ $ \cdots,\bm{k}_i\overline{\lambda}_i,\cdots)$.
To extract spin entanglement, 
the momenta of all nucleons are fixed by first projecting $\ket{\overline{\Phi}_{1\cdots N}}$ with the operator $\mathcal{P}(\bm{p}_1,\cdots,\bm{p}_N)=\frac{1}{N!}\sum_{\lambda_{1}\cdots \lambda_{N}}\ket{\bm{p}_1\lambda_{1},\cdots,\bm{p}_N\lambda_{N}}\langle\bm{p}_1\lambda_{1},\cdots,$ $\bm{p}_N\lambda_{N}|$ $(\lambda_{1},\cdots,\lambda_{N}=\pm\frac{1}{2})$,
after which $\ket{\overline{\Phi}_{1\cdots N}}$ becomes
\begin{align}
&\mathcal{P}(\bm{p}_1,\cdots,\bm{p}_N)\ket{\overline{\Phi}_{1\cdots N}}\nonumber\\
=&\sum_{\lambda_{1}\cdots \lambda_{N}}\mathcal{B}(\bm{p}_1\lambda_{1},\cdots,\bm{p}_N\lambda_{N})\ket{\bm{p}_1\lambda_{1},\cdots,\bm{p}_N\lambda_{N}}\nonumber\\
\to&\sum_{\lambda_{1}\cdots \lambda_{N}}\mathcal{B}(\bm{p}_1\lambda_1,\cdots,\bm{p}_N\lambda_{N})\ket{\tfrac{1}{2}-\lambda_{1},\cdots,\tfrac{1}{2}-\lambda_{N}}\nonumber\\
\equiv& \ket{\Phi_{1\cdots N}},
\label{N_Qubit_Projected}
\end{align}
which could be mapped to a $N$-qubit state as shown in Eq.~\eqref{N_Qubit_Projected}.
For the multinucleon systems with $N\leq 4$,
a proof-of-concept experimental setup to fix the nucleon momenta is given in Fig.~\ref{TMF},
where two- and three-nucleon states are taken as explicit examples.
Here, a destructive momentum measurement should be avoided, if it destroys the spin entanglement of nucleons.
 Instead, a number of narrow slits are arranged to select outgoing nucleons in specific directions.
 Experimentally, the total energy $E_N$ and the total momentum $\bm{P}_N$ of the multineutron can often be determined.
 For example, in the $^{3}\text{H}(t,{}^3\text{He})\,{}^3 n$ reaction adopted by Ref.~\cite{Miki:2023grx},
 the total energy and the total momentum of the trineutron ${}^3n$ can be deduced from those of the incoming triton
 and the outgoing ${}^3$He.
 As the momentum directions of all the nucleons are known, the unknown momentum magnitudes can be determined from $E_N$ and $\bm{P}_N$ by energy and momentum conservation laws.
 It is straightforward to see that
 this method works for multinucleons with $N\leq4$.
   
As shown in Sec.~\ref{EM}, the reduced density matrices  
play a fundamental role in constructing concurrence, GMC, GBC, and CF. 
Explicitly,
if $\rho_1$ is known for the two-qubit system, the concurrence $\mathcal{C}_{1|2}(\ket{\Phi_{12}})$ can be constructed from Eq.~\eqref{concurrence2};
if $\rho_1$, $\rho_2$, and $\rho_3$ are known for the three-qubit system, $\mathcal{C}_\text{GMC}(\ket{\Phi_{123}})$, $\mathcal{C}_\text{GBC}(\ket{\Phi_{123}})$, and $\mathcal{C}_\text{CF}(\ket{\Phi_{123}})$ can be constructed from bipartite concurrences 
$\mathcal{C}_{1|23}$, $\mathcal{C}_{2|13}$,  and $\mathcal{C}_{3|12}$ according to Eqs.~\eqref{CGMC3}--\eqref{CCF3}.
In the next subsection, we will discuss how to determine the reduced density matrices for multinucleons in polarization experiments \cite{Wolfenstein:1956,MacGregor:1960,Hoshizaki:1968,Ohlsen:1972}.

\begin{figure}

\centering
  \includegraphics[width=0.8\linewidth]{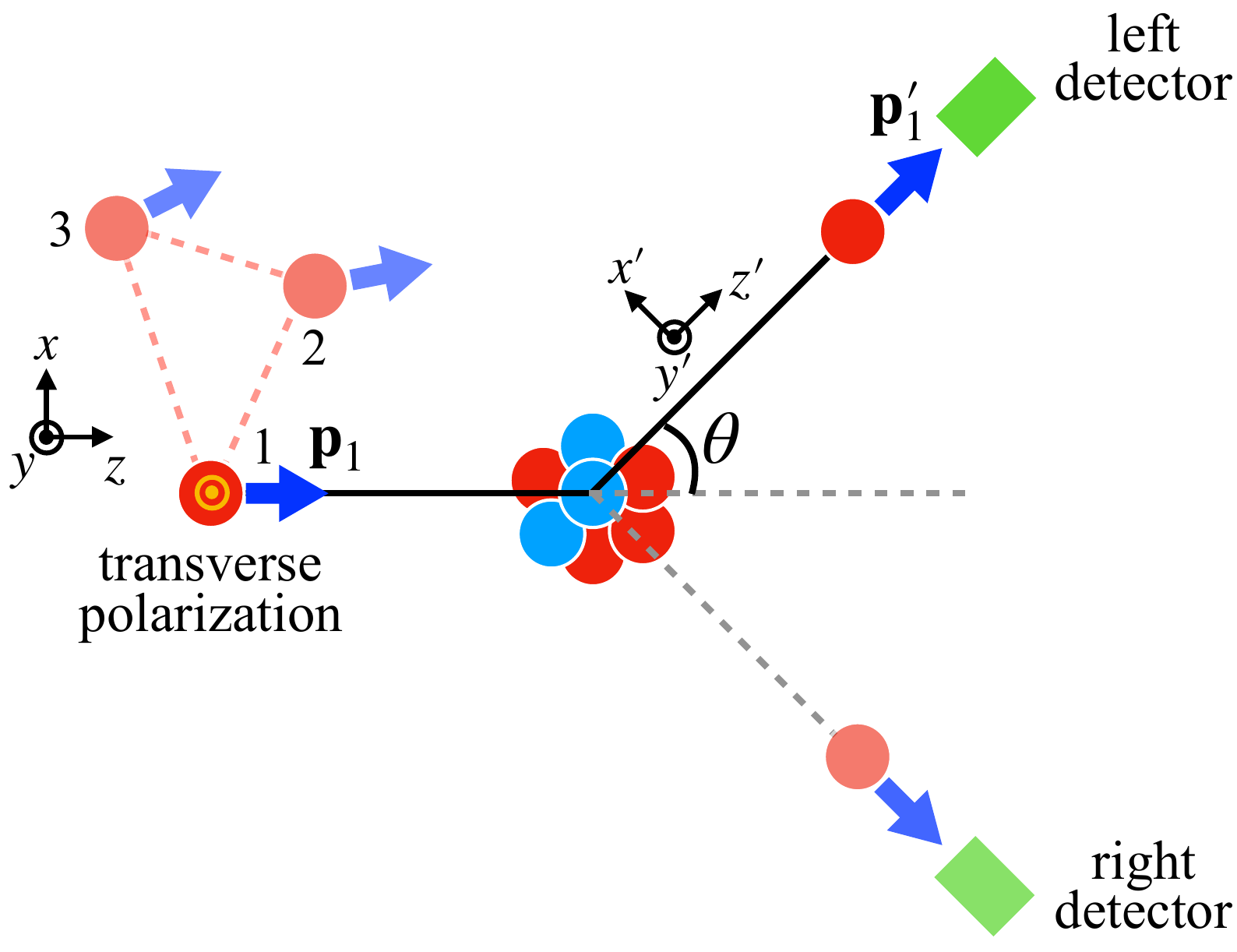}
  \caption{The experimental setup for determining the transverse polarization $\braket{\sigma_{1y}}=\text{tr}(\sigma_{1y}\ket{\Phi_{123}}\bra{\Phi_{123}})$ of the three-nucleon state by scattering the first nucleon with the spin-0 nucleus. Two helicity frames are adopted, with the $z$ and $z'$ axes along the directions of $\bm{p}_1$ and $\bm{p}'_1$, the $y$ and $y'$ axes along the directions of $\bm{p}_1\times\bm{p}_1'$,
  and the $x$ and $x'$ axes chosen to form right-handed coordinate systems. 
The transverse polarizations of the other nucleons
  can be determined in a similar way.
  }

  \label{SS}
       
\end{figure}

\begin{figure}

\centering
  \includegraphics[width=\linewidth]{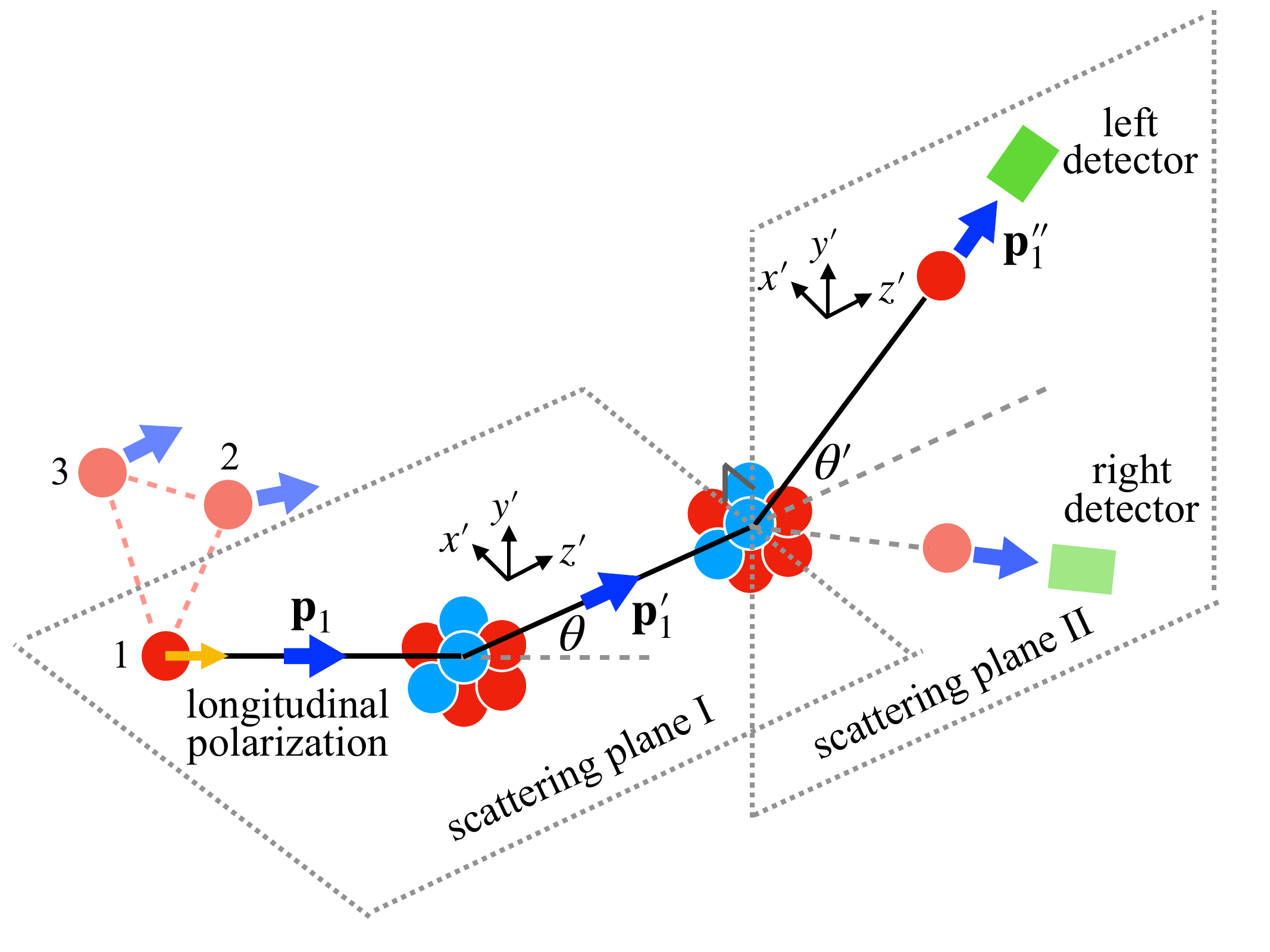}
  \caption{
  The experimental setup for determining the longitudinal polarization $\braket{\sigma_{1z}}=\text{tr}(\sigma_{1z}\ket{\Phi_{123}}\bra{\Phi_{123}})$ of the three-nucleon state by scattering the first nucleon with two spin-0 nuclei sequentially, with the first scattering plane spanned by $\bm{p}_1$ and $\bm{p}_1'$ perpendicular to the second one spanned by $\bm{p}_1'$ and $\bm{p}_1''$. 
  The longitudinal polarizations of the other nucleons
  can be determined in a similar way.
  }

  \label{DS}
       
\end{figure}

\subsection{Nuclear State Tomography}

As shown before, 
the reduced density matrix $\rho_1=\text{tr}_2(\ket{\Phi_{12}}$ $\bra{\Phi_{12}})$ is
the building block for the concurrence $\mathcal{C}_{1|2}(\ket{\Phi_{12}})$ of the two-nucleon state $\ket{\Phi_{12}}$,
 while the reduced density matrices $\rho_1=\text{tr}_{23}(\ket{\Phi_{123}}\bra{\Psi_{123}})$,
 $\rho_2=\text{tr}_{13}(\ket{\Phi_{123}}\bra{\Phi_{123}})$,
 and $\rho_3=\text{tr}_{12}($ $\ket{\Phi_{123}}\bra{\Phi_{123}})$ 
 are the building blocks for
 the GME measures of the three-nucleon state $\ket{\Phi_{123}}$.
 Consider a general reduced density matrix $\rho_i$ for the $i$th nucleon, which is parametrized by
 \begin{align}
 \rho_i=\frac{1}{2}(1+\braket{\sigma_{ix}}\sigma_{ix}+\braket{\sigma_{iy}}\sigma_{iy}+\braket{\sigma_{iz}}\sigma_{iz}),
 \end{align}
 with $\braket{\sigma_{i(x,y,z)}}=\text{tr}[\sigma_{i(x,y,z)}\ket{\Phi_{123}}\bra{\Phi_{123}}]=\text{tr}_{1}[\sigma_{i(x,y,z)}\rho_i]$ being the three spin polarizations of the $i$th nucleon along the $x$, $y$, $z$ axes. For the two-nucleon state $\ket{\Phi_{12}}$, one has
 \begin{align}
 \mathcal{C}_{1|2}(\ket{\Phi_{12}})=\sqrt{1-\braket{\sigma_{1x}}^2-\braket{\sigma_{1y}}^2-\braket{\sigma_{1z}}^2},
 \label{Concurrence2}
 \end{align}
 while for the three-nucleon state $\ket{\Phi_{123}}$, one has
 \begin{align}
 \mathcal{C}_{i|jk}(\ket{\Phi_{123}})=\sqrt{1-\braket{\sigma_{ix}}^2-\braket{\sigma_{iy}}^2-\braket{\sigma_{iz}}^2},
  \label{Concurrence3}
 \end{align}
 with $i,j,k=1,2,3$. As shown in Eqs.~\eqref{CGMC3}--\eqref{CCF3},
 $\mathcal{C}_{1|23}(\ket{\Phi_{123}})$, $\mathcal{C}_{2|13}(\ket{\Phi_{123}})$, and $\mathcal{C}_{3|12}(\ket{\Phi_{123}})$ constitute the whole set of building blocks to construct GME measures such as GMC, GBC, and CF for the three-nucleon state.
 As a result, the concurrence is determined for the two-nucleon state if the three spin polarizations are measured for one of the two nucleons,
 and the GME measures are determined for the three-nucleon state if the spin polarizations are measured for all of the three nucleons.
 For the $i$th nucleon, the three spin polarizations can be divided into two classes, two transverse polarizations perpendicular to the momentum direction plus one longitudinal polarization parallel to the momentum direction.
These two kinds of polarizations can be determined by the single- and double-scattering experiments shown in Figs.~\ref{SS} and \ref{DS}, respectively.
While it is the three-nucleon state that is taken as an example, the discussions can be easily adapted to the two-nucleon state.

In Fig.~\ref{SS}, the initial state is given by $\ket{\Phi_{123}}$, the three-nucleon state after momentum fixing. 
To determine the transverse polarization $\braket{\sigma_{1y}}$,
it is the first nucleon with the momentum $\bm{p}_1$ that is taken to scatter with the spin-0 nucleus, while the other two nucleons are left undetected.
$\braket{\sigma_{1y}}$ is then related to the so-called left-right asymmetry by \cite{Ohlsen:1972,Bai:2023hrz}
\begin{align}
\braket{\sigma_{1y}}=\frac{1}{\mathcal{A}(\theta)}\frac{N_L(\theta)-N_R(\theta)}{N_L(\theta)+N_R(\theta)},
\end{align}
where $\mathcal{A}(\theta)$ is the analyzing power which could be measured for the nucleon-nucleus system priorly,
$N_L(\theta)$ and $N_R(\theta)$ are the number of events counted by the left and right detectors,
and $\theta$ is the relative angle between $\bm{p}_1$ and $\bm{p}_1'$.
The other transverse polarization $\braket{\sigma_{1x}}$ of the first nucleon could be determined by rotating the scattering plane in such a way that $\bm{p}_1\times\bm{p}_1'$ is parallel to the $x$ axis.
The transverse polarizations of the other two nucleons could be determined in the same way, and they are also needed to construct GME measures for the three-nucleon states.

In Fig.~\ref{DS}, the longitudinal polarization $\braket{\sigma_{1z}}$ is measured by scattering the first nucleon with two spin-0 nuclei sequentially.
Here, the scattering plane I spanned by $\bm{p}_1$ and $\bm{p}_1'$ is perpendicular to the scattering plane II spanned by $\bm{p}_1'$ and $\bm{p}_1''$, with the left and right detectors placed in the scattering plane II.
The helicity frame of the first nucleon after the first scattering is also shown explicitly in the figure,
with the $z'$ axis aligned with $\bm{p}'_1$, the $y'$ axis perpendicular to the scattering plane I, and the $x'$ axis chosen to form a right-handed coordinate system. $\braket{\sigma_{1z}}$ could then be given by \cite{Ohlsen:1972,Bai:2023hrz}
\begin{align}
&\braket{\sigma_{1z}}=\frac{1}{\mathcal{K}_{z}^{x'}(\theta)}\{\braket{\sigma_{1x'}}[1+\braket{\sigma_{1y}}\mathcal{A}(\theta)]-\braket{\sigma_{1x}}\mathcal{K}_x^{x'}(\theta)\},\\
&\braket{\sigma_{1x'}}=\frac{1}{\mathcal{A}'(\theta')}\frac{N_L'(\theta')-N_R'(\theta')}{N_L'(\theta')+N_R'(\theta')}.
\end{align}
Here, $\theta$ and $\theta'$ are the scattering angles for the first and second scattering,
$\mathcal{A}(\theta)$ and $\mathcal{A}'(\theta')$ are the analyzing powers of the first and second scattering,
$\mathcal{K}_x^{x'}(\theta)$ and $\mathcal{K}_z^{x'}(\theta)$ are the polarization transfer coefficients,
and $N_L'(\theta')$ and $N_R'(\theta')$ are the numbers of events counted by the left and right detectors.
If the same spin-0 nucleus is chosen for the two scattering processes, the two analyzing powers are related by $\mathcal{A}'(\theta')=\mathcal{A}(\theta')$.
 $\mathcal{A}(\theta)$, $\mathcal{A}'(\theta')$, $\mathcal{K}_x^{x'}(\theta)$, and $\mathcal{K}_z^{x'}(\theta)$ can be determined by using the standard polarization experiments,
 while
the transverse polarizations $\braket{\sigma_{1x}}$ and $\braket{\sigma_{1y}}$ should be measured in prior by following Fig.~\ref{SS}.

Before applying the above discussions to multiprotons and multineutrons, two subtleties should be thought about carefully.
Considering the detection efficiency,
multiprotons are apparently better targets than multineutrons,
as it is much easier to detect protons than neutrons in nuclear experiments.
However, the final-state Coulomb interactions between protons have a sizable impact on the late-time evolution of multiprotons,
and it is still unclear whether this would cause any trouble for interpreting the measured spin entanglement.
Nevertheless, these subtleties are by no means insurmountable difficulties and should not change the above discussions essentially.

\subsection{Upper Bounds for Spin Entanglement}

In general, double-scattering experiments are more difficult than single-scattering ones. Let's consider the extreme situation where only the transverse polarizations $\braket{\sigma_{ix}}$ and $\braket{\sigma_{iy}}$ are measured. 
Thus, one does not have enough data to derive concurrence and GME measures.
However, even in this case, one can get useful information on spin entanglement of multinucleons.
Thanks to Eqs.~\eqref{Concurrence2} and \eqref{Concurrence3},
it is straightforward to see that 
$\mathcal{C}_{1|2}(\ket{\Phi_{12}})\leq\sqrt{1-\braket{\sigma_{1x}}^2-\braket{\sigma_{1y}}^2}$
and $\mathcal{C}_{i|jk}(\ket{\Phi_{123}})\leq\sqrt{1-\braket{\sigma_{ix}}^2-\braket{\sigma_{iy}}^2}$.
In other words, $\braket{\sigma_{ix}}$ and $\braket{\sigma_{iy}}$ alone give upper bounds to concurrences.
Then, according to Eq.~\eqref{CGMC3}, GMC, one of the GME measures, satisfies $\mathcal{C}_\text{GMC}(\ket{\Phi_{123}})\leq\text{min}\{$ $\sqrt{1-\braket{\sigma_{ix}}^2-\braket{\sigma_{iy}}^2}\}$, with $i=1,2,3$.
Therefore, they also give the upper bound for the GMC of the three-nucleon state.

\section{Conclusions}
\label{Concl}

Multinucleons made of identical nucleons can appear naturally in multinucleon emissions from unstable nuclei and knockout reactions.
We study the problem of experimental determination of the amount of their spin entanglement,
with special emphasis on two- and three-nucleon systems.
It is shown that the concurrence and the GME measures for the two- and three-nucleon systems
can all be determined by measuring spin polarizations of nucleons,
with the possible experimental setups discussed in detail.
In principle, a generalization to the four-nucleon system is also possible.
However, this would need the measurement of spin correlation functions of multiple pairs of nucleons in single- and double-scattering processes,
which is much more difficult than the two- and three-nucleon cases. 

There are several open directions related to our work.
The first one is to determine the effects of final-state interactions on spin entanglement,
which can help to develop a better understanding of the measured concurrence and GME measures in the case of multiprotons.
The second one is to optimize the NST setups such that it can be adapted to realistic experimental situations.
The present work should be understood more or less as a proof of concept, and it is expected that there is some room for further improvements.
The third one is to figure out experimental methods to probe entanglement inside bound nuclei.
The present work is devoted to totally unbound multinucleons where all the nucleons eventually move apart to the infinity.
Generalizing it to bound nuclei can increase the experimental prospects of nuclear entanglement significantly.
These problems will be studied in future publications.

\section*{Acknowledgments}
This work is supported by the National Natural Science Foundation of China (Grants No.\ 12375122, 
No.\ 12035011, No.\ 11975167, and No.\ 12147101), 
the National Key R\&D Program of China (Contract No.\ 2023YFA1606503),
and the Fundamental Research Funds for the Central Universities (Grant No.\ B230201022).

\end{document}